\documentclass[12pt,draftclsnofoot,onecolumn,peerreview]{IEEEtran}

\newtheorem{Thm}{Theorem}

\usepackage{amsmath}
\usepackage{amssymb}
\usepackage{psfrag}
\usepackage{subfigure}
\usepackage{graphicx}

\title{On Coding for Reliable Communication \\ over Packet Networks}

\author{Desmond S. Lun,
Muriel M\'edard,
Ralf Koetter, 
and Michelle Effros
\thanks{This work was supported by 
the National Science Foundation under 
grant nos.\  CCR-0093349, CCR-0325496, and CCR-0325673; 
by the Army Research Office through University of California
subaward no.\  S0176938; and
by the Office of Naval Research under 
grant no.\  N00014-05-1-0197.} 
\thanks{This paper was presented in part at the 
42nd Annual Allerton Conference on Communication, Control, and
Computing, Monticello, IL, September--October 2004;
and in part 
at the 2005 International Symposium on Information Theory, Adelaide,
Australia, September 2005.}
\thanks{D. S. Lun and M. M\'edard
are with the Laboratory for
Information and Decision Systems, Massachusetts Institute of
Technology, Cambridge, MA 02139, USA (e-mail: dslun@mit.edu,
medard@mit.edu).}
\thanks{R. Koetter 
is with the Coordinated Science Laboratory,
University of Illinois at Urbana-Champaign, 
Urbana, IL 61801, USA (e-mail: koetter@uiuc.edu).}
\thanks{M. Effros is with the Department of Electrical Engineering,
California Institute of Technology, Pasadena, CA 91125, USA 
(e-mail: effros@caltech.edu).}}

\begin{document}
\maketitle

\begin{abstract}

We present a capacity-achieving coding scheme for unicast or multicast
over lossy packet networks.  In the scheme, intermediate nodes perform
additional coding yet do not decode nor even wait for a block of packets
before sending out coded packets.  Rather, whenever they have a
transmission opportunity, they send out coded packets formed from random
linear combinations of previously received packets.  All coding and
decoding operations have polynomial complexity.

We show that the scheme is capacity-achieving as long as packets
received on a link arrive according to a process that has an average
rate.  Thus, packet losses on a link may exhibit correlation in time or
with losses on other links.  In the special case of Poisson traffic with
i.i.d.\ losses, we give error exponents that quantify the rate of decay
of the probability of error with coding delay.  Our analysis of the
scheme shows that it is not only capacity-achieving, but that the
propagation of packets carrying ``innovative'' information follows the
propagation of jobs through a queueing network, and therefore fluid flow
models yield good approximations.  We consider networks with both lossy
point-to-point and broadcast links, allowing us to model both wireline
and wireless packet networks.  

\end{abstract}

\IEEEpeerreviewmaketitle

\section{Introduction}

Network information theory generally focuses on applications that,
in the open systems interconnection (OSI) model of network architecture,
lie in the physical layer.  In this context, there are some
networked systems, such as those represented by the multiple-access
channel and the broadcast channel, that are well understood, but there
are many that remain largely intractable.  Even some very simple
networked systems, such as those represented by the relay channel and
the interference channel, have unknown capacities.

But the relevance of network information theory is not limited to the
physical layer.  In practice, the physical layer never provides a
fully-reliable bit pipe to higher layers, and reliability then falls on
the data link control, network, and transport layers.  These layers need
to provide reliability not only because of an unreliable physical layer,
but also because of packet losses resulting from causes such as
congestion (which leads to buffer overflows) and interference (which
leads to collisions).  Rather than coding over channel symbols, though,
coding is applied over packets, i.e.\  rather than determining each
node's outgoing channel symbols through arbitrary, causal mappings of
their received symbols, the contents of each node's outgoing packets are
determined through arbitrary, causal mappings of the contents of their
received packets.  Such packet-level coding offers an alternative domain
for network information theory and an alternative opportunity for
efficiency gains resulting from cooperation, and it is the subject of
our paper.

Packet-level coding differs from symbol-level coding in three principal
ways:  First, in most packetized systems, packets received in error are
dropped, so we need to code only for resilience against erasures and not
for noise.  Second, it is acceptable to append a degree of
side-information to packets by including it in their headers.  Third,
packet transmissions are not synchronized in the way that symbol
transmissions are---in particular, it is not reasonable to assume that
packet transmissions occur on every link in a network at identical,
regular intervals.  These factors make for a different, but related,
problem to symbol-level coding.  Thus, our work addresses a problem of
importance in its own right as well as possibly having implications to
network information theory in its regular, symbol-level setting.

Aside from these three principal differences, packet-level coding is
simply symbol-level coding with packets as the symbols.  Thus, given a
specification of network use (i.e.\  packet injection times), a code
specifies the causal mappings that nodes apply to packets to determine
their contents; and, given a specification of erasure locations in
addition to the specification of network use (or, simply, given packet
reception times corresponding to certain injection times), we can define
capacity as the maximum reliable rate (in packets per unit time) that
can be achieved.  Thus, when we speak of capacity, we speak of Shannon
capacity as it is normally defined in network information theory (save
with packets as the symbols).  We do not speak of the various other
notions of capacity in networking literature.

The prevailing approach to packet-level coding uses a feedback code:
Automatic repeat request (ARQ) is used to request the retransmission of
lost packets either on a link-by-link basis, an end-to-end basis, or
both.  This approach often works well and has a sound theoretical basis:
It is well known that, given perfect feedback, retransmission of lost
packets is a capacity-achieving strategy for reliability on a
point-to-point link (see, for example, \cite[Section 8.1.5]{cot91}).  
Thus, if achieving a network connection meant
transmitting packets over a series of uncongested point-to-point links
with reliable, delay-free feedback, then retransmission is clearly
optimal.  This situation is approximated in lightly-congested,
highly-reliable wireline networks, but it is generally not the case.
First, feedback may be unreliable or too slow, which is often the
case in satellite or wireless networks or when servicing real-time
applications.  Second, congestion can always arise in packet networks;
hence the need for retransmission on an end-to-end basis.  But, if the
links are unreliable enough to also require retransmission on a link-by-link
basis, then the two feedback loops can interact in complicated, and
sometimes undesirable, ways \cite{liu05, lim02}.  Moreover, such end-to-end
retransmission requests are not well-suited for multicast connections,
where, because requests are sent by each terminal as packets are
lost, there may be many requests,
placing an unnecessary load on the network
and possibly overwhelming the source; and packets that are retransmitted
are often only of use to a subset of the terminals and therefore
redundant to the remainder.
Third, we may not be dealing with point-to-point links at all. 
Wireless networks are the obvious case in point.
Wireless links are often treated as
point-to-point links, with packets being routed hop-by-hop toward their
destinations, but, if the lossiness of the medium is accounted for, this
approach is sub-optimal.  In general, the broadcast nature of the links
should be exploited; and, in this case, a great deal of feedback would
be required to achieve reliable communication using a
retransmission-based scheme.

In this paper, therefore, we eschew this approach in favor of one that
operates mainly in a feedforward manner.  Specifically, we consider the
following coding scheme:  Nodes store the packets they receive into
their memories and, whenever they have a transmission opportunity, they
form coded packets with random linear combinations of their memory
contents.  This strategy, we shall show, is capacity-achieving, for both
single unicast and single multicast connections and for models of both
wireline and wireless networks, as long as packets received on each link
arrive according to a process that has an average rate.  Thus, packet
losses on a link may exhibit correlation in time or with losses on other
links, capturing various mechanisms for loss---including
collisions.

The scheme has several other attractive properties:  It is
decentralized, requiring no coordination among nodes; and it can be
operated ratelessly, i.e.\  it can be run indefinitely until successful
decoding (at which stage that fact is signaled to other nodes,
requiring an amount of feedback that, compared to ARQ, is small), which
is a particularly useful property in packet networks, where loss rates
are often time-varying and not known precisely.  

Decoding can be done by matrix inversion, which is a polynomial-time
procedure.  Thus, though we speak of random coding, our work differs
significantly from that of Shannon~\cite{sha48a, sha48b} and
Gallager~\cite{gal65} in that we do not seek to demonstrate existence.
Indeed, the existence of capacity-achieving linear codes for the
scenarios we consider already follows from the results of~\cite{dgp06}.
Rather, we seek to show the asymptotic rate optimality of a specific
scheme that we believe may be practicable and that can be considered as
the prototype for a family of related, improved schemes; for example,  LT
codes~\cite{lub02}, Raptor codes~\cite{sho06}, Online
codes~\cite{may02}, RT oblivious erasure-correcting codes~\cite{bds04},
and the greedy random scheme proposed in~\cite{pfs05} are related coding
schemes that apply only to specific, special networks
but, using varying degrees of feedback,
achieve lower decoding complexity or memory usage.  Our work
therefore brings forth a natural code design problem, namely to find
such related, improved schemes.

We begin by describing the coding scheme in the following section.  In
Section~\ref{sec:model}, we describe our model and illustrate it with
several examples.  In Section~\ref{sec:coding_theorems}, we present
coding theorems that prove that the scheme is capacity-achieving and, in
Section~\ref{sec:error_exponents}, we strengthen these results in the
special case of Poisson traffic with i.i.d.\  losses by giving error
exponents.  These error exponents allow us to quantify the rate of decay
of the probability of error with coding delay and to determine the
parameters of importance in this decay.

\section{Coding scheme}
\label{sec:coding_scheme}

We suppose that, at the source node, we have $K$ message packets
$w_1, w_2, \ldots, w_K$, which are vectors of length $\lambda$ over the
finite field $\mathbb{F}_q$.  (If the packet length is $b$ bits, then we
take $\lambda = \lceil b / \log_2 q \rceil$.)  The message packets are
initially present in the memory of the source node. 

The coding operation performed by each node is simple to describe and is
the same for every node:  Received packets are stored into the node's
memory, and packets are formed for injection with random
linear combinations of its memory contents 
whenever a packet injection occurs on an
outgoing link.  The coefficients of the combination are drawn uniformly
from $\mathbb{F}_q$.  

Since all coding is linear, we can write any
packet $x$ in the network as a linear combination of 
$w_1, w_2, \ldots, w_K$, namely,
$x = \sum_{k=1}^K \gamma_k w_k$. 
We call $\gamma$ the \emph{global
encoding vector} of $x$, and we assume that it is sent along with $x$,
as side information in its header. 
The overhead this incurs (namely, $K \log_2 q$ bits)
is negligible if packets are sufficiently large.

Nodes are assumed to have unlimited memory.  The scheme can be modified
so that received packets are stored into memory only if their global
encoding vectors are linearly-independent of those already stored.  This
modification keeps our results unchanged while ensuring that nodes never
need to store more than $K$ packets.

A sink node collects packets and, if it has $K$ packets with
linearly-independent global encoding vectors, it is able to recover the
message packets.  Decoding can be done by Gaussian elimination.  The
scheme can be run either for a predetermined duration or, in the case
of rateless operation, until successful decoding at the sink nodes.  We
summarize the scheme in Figure~\ref{fig:summary_RLC}.

\begin{figure}
\centering
\framebox{
\begin{minipage}{0.88\textwidth}
\noindent\textbf{Initialization:}
\begin{itemize}
\item The source node stores the message packets $w_1, w_2, \ldots, w_K$ in its memory.
\end{itemize}
\noindent\textbf{Operation:}
\begin{itemize}
\item When a packet is received by a node,
\begin{itemize}
\item the node stores the packet in its memory.
\end{itemize}
\item When a packet injection occurs on an outgoing link of a node,
\begin{itemize}
\item the node forms the packet from a random linear combination of 
the packets in its memory.  Suppose the node has $L$ packets $y_1, y_2,
\ldots, y_L$ in its memory.  Then the packet formed is
\[
x := \sum_{l=1}^L \alpha_l y_l,
\]
where $\alpha_l$ is chosen according to a uniform distribution over the
elements of $\mathbb{F}_q$.  
The packet's global encoding vector $\gamma$, which satisfies
$x = \sum_{k=1}^K \gamma_k w_k$, is placed in its header.
\end{itemize}
\end{itemize}
\noindent\textbf{Decoding:}
\begin{itemize}
\item Each sink node performs Gaussian elimination on the set of global
encoding vectors from the packets in its memory.  If it is able to find
an inverse, it applies the inverse to the packets to obtain $w_1,
w_2, \ldots, w_K$; otherwise, a decoding error occurs.
\end{itemize}
\end{minipage} }
\caption{Summary of the random linear coding scheme we consider.}
\label{fig:summary_RLC}
\end{figure}

The scheme is carried out for a single block of $K$ message packets at
the source.  If the source has more packets to send, then the scheme is
repeated with all nodes flushed of their memory contents.

Similar random linear coding schemes are described in \cite{hkm03, ho04,
hmk06, cwj03} for the application of multicast over lossless wireline
packet networks, in \cite{dmc06} for data dissemination, in \cite{adm05}
for data storage, and in \cite{gkr05} for content distribution over
peer-to-peer overlay networks.  Other coding schemes for lossy packet
networks are described in \cite{dgp06} and \cite{khs05-multirelay}; the
scheme described in the former requires placing in the packet headers
side information that grows with the size of the network, while that
described in the latter requires no side information at all, but
achieves lower rates in general.  Both of these coding schemes,
moreover, operate in a block-by-block manner, where coded packets are
sent by intermediate nodes only after decoding a block of received
packets---a strategy that generally incurs more delay than the scheme we
consider, where intermediate nodes perform additional coding yet do not
decode \cite{pfs05}.  

\section{Model}
\label{sec:model}

Existing models used in network information theory (see, for example,
\cite[Section 14.10]{cot91}) are generally conceived for symbol-level
coding and, given the peculiarities of packet-level coding, are not
suitable for our purpose.  One key difference, as we mentioned, is that
packet transmissions are not synchronized in the way that symbol
transmissions are.  Thus, we do not have a slotted system where packets
are injected on every link at every slot, and we must therefore have a
schedule that determines when (in continuous time)
and where (i.e.\  on which link) each packets is injected.  In this
paper, we assume that such a schedule is given, and we do not address
the problem of determining it.  This problem, of determining the
schedule to use, is a difficult problem in its own right, especially in
wireless packet networks.  Various instances of the problem are treated
in \cite{lrm06, sae05, wck06, wck05, wcz05, xiy05, xiy06}.

Given a schedule of packet injections, the network responds with packet
receptions at certain nodes.  The difference between wireline and
wireless packet networks, in our model, is that the reception of any
particular packet may only occur at a single node in wireline packet
networks while, in wireless packet networks, it may occur at more than
one node.

The model, which we now formally describe, is one that we believe is an
accurate abstraction of packet networks as they are viewed at the level
of packets, given a schedule of packet injections.  In particular, our
model captures various phenomena that complicate the efficient operation
of wireless packet networks, including interference (insofar as it is
manifested as lost packets, i.e.\  as collisions), fading (again,
insofar as it is manifested as lost packets), and the broadcast nature
of the medium.

We begin with wireline packet networks.  We model a wireline packet
network (or, rather, the portion of it devoted to the connection we wish
to establish) as a directed graph $\mathcal{G} =
(\mathcal{N},\mathcal{A})$, where $\mathcal{N}$ is the set of nodes and
$\mathcal{A}$ is the set of arcs.  Each arc $(i,j)$ represents a lossy
point-to-point link.  Some subset of the packets injected into arc
$(i,j)$ by node $i$ are lost; the rest are received by node $j$ without
error.  We denote by $z_{ij}$ the average rate at which packets are
received on arc $(i,j)$.  More precisely, suppose that the arrival of
received packets on arc $(i,j)$ is described by the counting process
$A_{ij}$, i.e.\  for $\tau \ge 0$, $A_{ij}(\tau)$ is the total number of
packets received between time 0 and time $\tau$ on arc $(i,j)$.  Then,
by assumption, $\lim_{\tau \rightarrow \infty} {A_{ij}(\tau)}/{\tau} =
z_{ij}$ a.s.  We define a lossy wireline packet network as a pair
$(\mathcal{G}, z)$.

We assume that links are delay-free in the sense that the arrival time
of a received packet corresponds to the time that it was injected into
the link.  Links with delay can be transformed into delay-free links in
the following way:  Suppose that arc $(i,j)$ represents a link with
delay.  The counting process $A_{ij}$ describes the arrival of received
packets on arc $(i,j)$, and we use the counting process $A_{ij}^\prime$
to describe the injection of these packets.  (Hence $A_{ij}^\prime$
counts a subset of the packets injected into arc $(i,j)$.)  We insert a
node $i^\prime$ into the network and transform arc $(i,j)$ into two arcs
$(i, i^\prime)$ and $(i^\prime, j)$.  These two arcs, $(i, i^\prime)$
and $(i^\prime, j)$, represent delay-free links where the arrival of
received packets are described by $A_{ij}^\prime$ and $A_{ij}$,
respectively.  We place the losses on arc $(i,j)$ onto arc $(i,
i^\prime)$, so arc $(i^\prime, j)$ is lossless and node $i^\prime$
simply functions as a first-in first-out queue.  It is clear that
functioning as a first-in first-out queue is an optimal coding strategy
for $i^\prime$ in terms of rate and complexity; hence, treating
$i^\prime$ as a node implementing the coding scheme of
Section~\ref{sec:coding_scheme} only deteriorates performance and is
adequate for deriving achievable connection rates.  Thus, we can
transform a link with delay and average packet reception rate $z_{ij}$
into two delay-free links in tandem with the same average packet
reception rate, and it will be evident that this transformation does not
change any of our conclusions.

For wireless packet networks, we model the network as a directed hypergraph 
$\mathcal{H} = (\mathcal{N},\mathcal{A})$, 
where $\mathcal{N}$ is the set of nodes and $\mathcal{A}$ is the set of
hyperarcs.  
A hypergraph is a generalization of a graph where generalized arcs,
called hyperarcs, connect two or more nodes.  Thus,
a hyperarc is a pair $(i,J)$, where $i$, the head, is an
element of $\mathcal{N}$, and $J$, the tail, is a non-empty subset of
$\mathcal{N}$.
Each hyperarc $(i,J)$ represents a lossy broadcast link.
For each $K \subset J$, some disjoint subset of the packets injected
into hyperarc $(i,J)$ by node $i$ are received by exactly the set of
nodes $K$ without error.

We denote by $z_{iJK}$ the average rate at which
packets, injected on hyperarc $(i,J)$, are received by exactly 
the set of nodes $K
\subset J$.  More precisely, suppose that the arrival of packets that
are injected on hyperarc $(i,J)$ and received by all nodes in $K$
(and no nodes in $\mathcal{N} \setminus K$) 
is described by the counting process $A_{iJK}$.  Then, by
assumption,
$\lim_{\tau \rightarrow \infty} {A_{iJK}(\tau)}/{\tau} = z_{iJK}$ 
a.s.
We define a lossy wireless packet network as a pair $(\mathcal{H}, z)$.

\subsection{Examples}

\subsubsection{Network of independent transmission lines with non-bursty
losses}

We begin with a simple example.  We consider a wireline
network where each transmission line experiences losses independently of
all other transmission lines, and the loss process on each line is
non-bursty, i.e.\  it is accurately described by an i.i.d.\  process.  

Consider the link corresponding to arc $(i,j)$.  Suppose the
loss rate on this link is $\varepsilon_{ij}$, i.e.\  packets are lost
independently with probability $\varepsilon_{ij}$.  Suppose further that
the injection of packets on arc $(i,j)$ is described by the counting
process $B_{ij}$ and has average rate $r_{ij}$, i.e.\  
$\lim_{\tau \rightarrow \infty} B_{ij}(\tau)/\tau = r_{ij}$ a.s.
The parameters $r_{ij}$ and $\varepsilon_{ij}$ are not necessarily 
independent and may well be functions of each other.

For the arrival of received packets, we have
\[
A_{ij}(\tau) = \sum_{k=1}^{B_{ij}(\tau)} X_k,
\]
where $\{X_k\}$ is a sequence of i.i.d.\  Bernoulli random variables
with $\Pr(X_k = 0) = \varepsilon_{ij}$.  Therefore
\[
\lim_{\tau \rightarrow \infty} \frac{A_{ij}(\tau)}{\tau}
= \lim_{\tau \rightarrow \infty} \frac{\sum_{k=1}^{B_{ij}(\tau)}
X_k}{\tau}
= \lim_{\tau \rightarrow \infty} \frac{\sum_{k=1}^{B_{ij}(\tau)}
X_k}{B_{ij}(\tau)} \frac{B_{ij}(\tau)}{\tau}
= (1-\varepsilon_{ij})r_{ij},
\]
which implies that 
\[
z_{ij} = (1-\varepsilon_{ij})r_{ij}.
\]

In particular, if the injection processes for all links are identical,
regular, deterministic processes with unit average rate 
(i.e.\  $B_{ij}(\tau) = 1 + \lfloor \tau \rfloor$ for all $(i,j)$),
then we recover the model frequently used in
information-theoretic analyses (for example, in \cite{dgp06,
khs05-multirelay}).

A particularly simple case arises when the injection processes are
Poisson.  In this case, $A_{ij}(\tau)$ and $B_{ij}(\tau)$ 
are Poisson random
variables with parameters $(1-\varepsilon_{ij})r_{ij}\tau$ and
$r_{ij}\tau$, respectively.  We shall revisit this case in
Section~\ref{sec:error_exponents}.

\subsubsection{Network of transmission lines with bursty losses}

We now consider a more complicated example, which attempts to model
bursty losses.  Bursty losses arise frequently in packet networks
because losses often result from phenomena that are time-correlated, for
example, fading and buffer overflows.  (We mention fading because a
point-to-point wireless link is, for our purposes, essentially
equivalent to a transmission line.)  In the latter case, losses are also
correlated across separate links---all links coming into a node
experiencing a buffer overflow will be subjected to losses.

To account for such correlations, Markov chains are often used.  Fading
channels, for example, are often modeled as finite-state Markov channels
\cite{wam95, gov96}, such as the Gilbert-Elliot channel \cite{mub89}.
In these models, a Markov chain is used to model the time evolution of
the channel state, which governs its quality.  Thus, if the channel is
in a bad state for some time, a burst of errors or losses is likely to
result.

We therefore associate with arc $(i,j)$ a continuous-time, irreducible
Markov chain whose state at time $\tau$ is $E_{ij}(\tau)$.  If
$E_{ij}(\tau) = k$, then the probability that a packet injected into
$(i,j)$ at time $\tau$ is lost is $\varepsilon_{ij}^{(k)}$.
Suppose that the steady-state probabilities of the chain are
$\{\pi_{ij}^{(k)}\}_k$.
Suppose further that
the injection of packets on arc $(i,j)$ is described by the counting
process $B_{ij}$ and that, conditioned on $E_{ij}(\tau) = k$, this
injection has average rate $r_{ij}^{(k)}$.
Then, we obtain
\[
z_{ij} = \pi_{ij}^\prime y_{ij}, 
\]
where
$\pi_{ij}$ and $y_{ij}$ denote the column vectors with
components $\{\pi_{ij}^{(k)}\}_k$ and $\{(1 - \varepsilon_{ij}^{(k)})
r_{ij}^{(k)}\}_k$,
respectively.  Our conclusions are not changed if the evolutions of the
Markov chains associated with separate arcs are correlated, such as would
arise from bursty losses resulting from buffer overflows.

If the injection processes are Poisson, then arrivals of received
packets are described by Markov-modulated Poisson processes
(see, for example, \cite{fim92}).

\subsubsection{Slotted Aloha wireless network}

We now move from wireline packet networks to wireless packet networks
or, more precisely, from networks of point-to-point links (transmission
lines) to networks where links may be broadcast links.

In wireless packet networks, one of most important issues is
medium access, 
i.e.\ determining how radio nodes share the wireless medium.  One simple,
yet popular, method for medium access control is slotted Aloha (see, for
example, \cite[Section 4.2]{beg92}), where nodes with packets to send
follow simple random rules to determine when they transmit.  In this
example, we consider a wireless packet network using slotted Aloha for
medium access control.
The example illustrates how a high degree of correlation in the loss
processes on separate links sometimes exists.

For the coding scheme we consider, nodes transmit whenever they are
given the opportunity and thus effectively always have packets to send.
So suppose that, in any given time slot, node $i$ transmits a packet on
hyperarc $(i,J)$ with probability $q_{iJ}$.  Let $p_{iJK|C}^\prime$ be
the probability that a packet transmitted on hyperarc $(i,J)$ is
received by exactly $K \subset J$ 
given that packets are transmitted on hyperarcs $C \subset \mathcal{A}$
in the same slot.
The distribution of
$p_{iJK|C}^\prime$ depends on many factors:  In the simplest case, if
two nodes close to each other transmit in the same time slot, then their
transmissions interfere destructively, resulting in a collision where
neither node's packet is received.  It is also possible that simultaneous
transmission does not necessarily result in collision, and 
one or more packets are received---sometimes referred to as
multipacket reception capability \cite{gvs88}.  It may even be the case
that physical-layer cooperative schemes, such as those presented in
\cite{kgg05, ltw04, aes04}, are used, where nodes that are not
transmitting packets are used to assist those that are.

Let $p_{iJK}$ be the unconditioned
probability that a packet transmitted on hyperarc $(i,J)$ is received by
exactly $K \subset J$.  So
\[
p_{iJK} = \sum_{C \subset \mathcal{A}} p_{iJK|C}^\prime
\left(\prod_{(j,L) \in C} q_{jL} \right)
\left(\prod_{(j,L) \in \mathcal{A}\setminus C} (1-q_{jL}) \right).
\]
Hence, assuming that time slots are of unit length, we see that
$A_{iJK}(\tau)$ follows a binomial distribution and
\[
z_{iJK} = q_{iJ} p_{iJK}.
\]

\begin{figure}
\begin{center}
\psfrag{1}[cc][cc]{\footnotesize 1}
\psfrag{2}[cc][cc]{\footnotesize 2}
\psfrag{3}[cc][cc]{\footnotesize 3}
\includegraphics[scale=1.25]{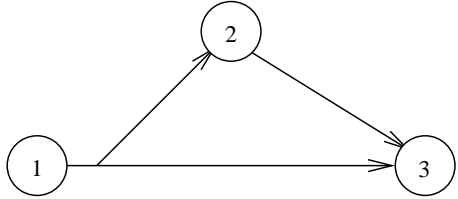}
\caption{The slotted Aloha relay channel.  We wish to establish a unicast
connection from node 1 to node 3.}
\label{fig:packet_relay}
\end{center}
\end{figure}

A particular network topology of interest is shown in
Figure~\ref{fig:packet_relay}.  The problem of setting up a unicast
connection from node 1 to node 3 in a slotted Aloha wireless network of
this topology is a problem that we refer to as the
slotted Aloha relay channel, in analogy to
the symbol-level relay channel widely-studied in network information
theory.  The latter problem is a well-known open problem, while the
former is, as we shall see, tractable and deals with the same issues of
broadcast and multiple access, albeit under different assumptions.

A case similar to that of slotted Aloha wireless networks is that of
untuned radio networks, which are detailed in \cite{prr05}.  
In such networks, nodes
are designed to be low-cost and low-power by sacrificing the ability for
accurate tuning of their carrier frequencies.
Thus, nodes transmit on random frequencies, which leads to random medium
access and contention.

\section{Coding theorems}
\label{sec:coding_theorems}

In this section, we specify achievable rate regions for the coding
scheme in various scenarios.  The fact that the regions we specify are
the largest possible (i.e.\  that the scheme is capacity-achieving) can
be seen by simply noting that the rate between any source and any sink
must be limited by the rate at which distinct packets are received
over any cut between that source and that sink.  A formal converse can
be obtained using the cut-set bound for multi-terminal networks (see
\cite[Section 14.10]{cot91}).

\subsection{Wireline networks}

\subsubsection{Unicast connections}
\label{sec:wireline_unicast}

\begin{figure}
\begin{center}
\psfrag{1}[cc][cc]{\footnotesize 1}
\psfrag{2}[cc][cc]{\footnotesize 2}
\psfrag{3}[cc][cc]{\footnotesize 3}
\includegraphics[scale=1.25]{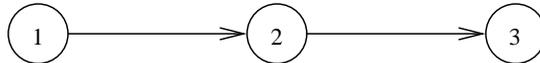}
\caption{A network consisting of two links in tandem.}
\label{fig:two_links}
\end{center}
\end{figure}

We develop our general result for unicast connections by extending
from some special cases.  We begin with the simplest non-trivial
case: that of two links in tandem (see Figure~\ref{fig:two_links}).

Suppose we wish to establish a connection of rate arbitrarily close to
$R$ packets per unit time from node 1 to node 3.  Suppose further
that the coding scheme is run for a total time $\Delta$, from time 0
until time $\Delta$, and that, in this time, a total of $N$ packets is
received by node 2.  We call these packets $v_1, v_2, \ldots, v_N$.

Any received packet $x$ in the network is a linear combination of $v_1,
v_2, \ldots, v_N$, so we can write 
\[
x = \sum_{n=1}^N \beta_n v_n.
\]
Since $v_n$ is formed by a random linear combination of the message
packets $w_1, w_2, \ldots, w_K$, we have
\[
v_n = \sum_{k=1}^K \alpha_{nk} w_k
\]
for $n = 1, 2, \ldots, N$, where each $\alpha_{nk}$ is drawn from a
uniform distribution over $\mathbb{F}_q$.  Hence
\[
x = \sum_{k=1}^K \left( \sum_{n=1}^N \beta_n \alpha_{nk} \right) w_k,
\]
and it follows that the $k$th component of the global encoding vector of
$x$ is given by
\[
\gamma_k = \sum_{n=1}^N \beta_n \alpha_{nk}.
\]
We call the vector $\beta$ associated with $x$ the \emph{auxiliary
encoding vector} of $x$, and we see that any node that receives 
$\lfloor K(1+\varepsilon) \rfloor$ or more packets with 
linearly-independent auxiliary encoding vectors has 
$\lfloor K(1+\varepsilon) \rfloor$ packets whose global encoding vectors
collectively form a random $\lfloor K(1+\varepsilon) \rfloor \times K$
matrix over $\mathbb{F}_q$, with all entries chosen uniformly.  If this
matrix has rank $K$, then node 3 is able to recover the message
packets.  The probability that a random 
$\lfloor K(1+\varepsilon) \rfloor \times K$ matrix has rank $K$ is, by a
simple counting argument,
$\prod_{k=1+\lfloor K(1+\varepsilon) \rfloor -K}^{\lfloor
K(1+\varepsilon) \rfloor} (1 - 1/q^k)$, which can be made arbitrarily
close to 1 by taking $K$ arbitrarily large.  Therefore, to determine
whether node 3 can recover the message packets, we essentially need only
to determine whether it receives $\lfloor K(1+\varepsilon) \rfloor$ or
more packets with linearly-independent auxiliary encoding vectors.

Our proof is based on tracking the propagation of what we call
\emph{innovative} packets.  Such packets are innovative in the sense
that they carry new, as yet unknown, information about $v_1, v_2,
\ldots, v_N$ to a node.\footnote{Note that, although we are 
ultimately concerned with recovering $w_1, w_2, \ldots, w_K$ rather
than $v_1, v_2, \ldots, v_N$, we define packets to be innovative with
respect to $v_1, v_2, \ldots, v_N$.  This serves to simplify our proof.
In particular, it means that we do not need to very strict in our
tracking of the propagation of innovative packets since the
number of innovative packets required at the sink is only a
fraction of $N$.}
It turns out that the propagation of innovative
packets through a network follows 
the propagation of jobs through a queueing network, 
for which fluid flow models give good approximations.
We present the
following argument in terms of this fluid analogy and defer the formal
argument to Appendix~\ref{app:formal_two_link_tandem}.

Since the packets being received by node 2 are the packets $v_1, v_2,
\ldots, v_N$ themselves, it is clear that every packet being received by
node 2 is innovative.  Thus, innovative packets arrive at node 2 at a
rate of $z_{12}$, and this can be approximated by fluid flowing in at
rate $z_{12}$.  These innovative packets are stored in node 2's memory,
so the fluid that flows in is stored in a reservoir.

Packets, now, are being received by node 3 at a rate of $z_{23}$, but
whether these packets are innovative depends on the contents of node 2's
memory.  If node 2 has more information about $v_1, v_2, \ldots, v_N$
than node 3 does, then it is highly likely that new information will be
described to node 3 in the next packet that it receives.  Otherwise, if
node 2 and node 3 have the same degree of information about $v_1, v_2,
\ldots, v_N$, then packets received by node 3 cannot possibly be
innovative.  Thus, the situation is as though fluid flows into node 3's
reservoir at a rate of $z_{23}$, but the level of node 3's reservoir is
restricted from ever exceeding that of node 2's reservoir.  The level of
node 3's reservoir, which is ultimately what we are concerned with, 
can equivalently be determined by fluid flowing out
of node 2's reservoir at rate $z_{23}$.

\begin{figure}
\begin{center}
\psfrag{2}[cc][cc]{\footnotesize 2}
\psfrag{3}[cc][cc]{\footnotesize 3}
\psfrag{#z_12#}[cc][cc]{\footnotesize $z_{12}$}
\psfrag{#z_23#}[cc][cc]{\footnotesize $z_{23}$}
\includegraphics[scale=1.25]{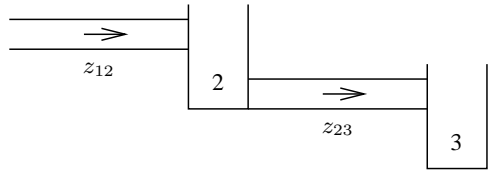}
\caption{Fluid flow system corresponding to two-link tandem network.}
\label{fig:two_pipes}
\end{center}
\end{figure}

We therefore see that the two-link tandem network in
Figure~\ref{fig:two_links} maps to the fluid flow system shown in
Figure~\ref{fig:two_pipes}.  It is clear that, in this system, fluid
flows into node 3's reservoir at rate $\min(z_{12}, z_{23})$.  This rate
determines the rate at which innovative packets---packets with new
information about $v_1, v_2, \ldots, v_N$ and, therefore, with
linearly-independent auxiliary encoding vectors---arrive at node 3.
Hence the time required for node 3
to receive $\lfloor K(1+\varepsilon) \rfloor$ packets with
linearly-independent auxiliary encoding vectors is, for large $K$,
approximately $K(1 + \varepsilon)/\min(z_{12}, z_{23})$, which implies
that a connection of rate arbitrarily close to $R$ packets per unit time
can be established provided that
\begin{equation}
R \le \min(z_{12}, z_{23}).
\label{eqn:130}
\end{equation}
Thus, we see that rate at which innovative packets are received by the
sink corresponds to an achievable rate.  Moreover,
the right-hand side of (\ref{eqn:130}) is indeed the capacity of the
two-link tandem network, and we therefore have the desired
result for this case.

\begin{figure}
\begin{center}
\psfrag{1}[cc][cc]{\footnotesize 1}
\psfrag{2}[cc][cc]{\footnotesize 2}
\psfrag{#cdots#}[cc][cc]{\footnotesize $\cdots$}
\psfrag{#L+1#}[cc][cc]{\footnotesize $L+1$}
\includegraphics[scale=1.25]{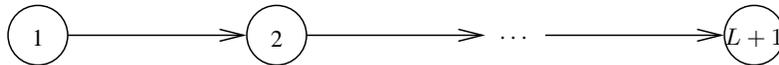}
\caption{A network consisting of $L$ links in tandem.}
\label{fig:l_links}
\end{center}
\end{figure}

We extend our result to another special case before considering
general unicast connections:  We consider the case of a tandem network
consisting of $L$ links and $L+1$ nodes (see Figure~\ref{fig:l_links}).

\begin{figure}
\begin{center}
\psfrag{1}[cc][cc]{\footnotesize 1}
\psfrag{2}[cc][cc]{\footnotesize 2}
\psfrag{#ddots#}[cc][cc]{\footnotesize $\ddots$}
\psfrag{#L+1#}[cc][cc]{\footnotesize $L+1$}
\psfrag{#z_12#}[cc][cc]{\footnotesize $z_{12}$}
\psfrag{#z_23#}[cc][cc]{\footnotesize $z_{23}$}
\psfrag{#z_L(L+1)#}[cc][cc]{\footnotesize $z_{L(L+1)}$}
\includegraphics[scale=1.25]{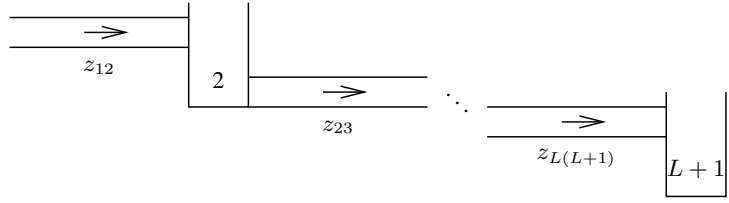}
\caption{Fluid flow system corresponding to $L$-link tandem network.}
\label{fig:l_pipes}
\end{center}
\end{figure}

This case is a straightforward extension of that of the two-link tandem
network.  It maps to the fluid flow system shown in
Figure~\ref{fig:l_pipes}.  In this system, it is clear that fluid flows 
into node $(L+1)$'s reservoir at rate 
$\min_{1 \le i \le L}\{z_{i(i+1)}\}$.  Hence a connection of rate
arbitrarily close to $R$ packets per unit time from node 1 to node $L+1$
can be established provided that
\begin{equation}
R \le \min_{1 \le i \le L}\{z_{i(i+1)}\}.
\label{eqn:150}
\end{equation}
Since the right-hand
side of (\ref{eqn:150}) is indeed the capacity of the $L$-link tandem
network, we therefore have the desired result for this
case.  A formal argument is in Appendix~\ref{app:formal_l_link_tandem}.

We now extend our result to general unicast connections.  The strategy
here is simple:  A general unicast connection can be formulated as a
flow, which can be decomposed into a finite number of paths.  Each of
these paths is a tandem network, which is the case that we have just
considered.

Suppose that we wish to establish a connection of rate arbitrarily close
to $R$ packets per unit time from source node $s$ to sink node $t$.
Suppose further that
\[
R \leq \min_{Q \in \mathcal{Q}(s,t)}
  \left\{\sum_{(i, j) \in \Gamma_+(Q)} z_{ij}
  \right\},
\]
where $\mathcal{Q}(s,t)$ is the set of all cuts between $s$ and $t$, and
$\Gamma_+(Q)$ denotes the set of forward arcs of the cut $Q$, i.e.\  
\[
\Gamma_+(Q) := \{(i, j) \in \mathcal{A} \,|\, i \in Q, j \notin Q\} .
\]
Therefore, by the max-flow/min-cut theorem (see, for example,
\cite[Section 3.1]{ber98}), there exists a
flow vector $f$ satisfying
\[
\sum_{\{j | (i,j) \in \mathcal{A}\}} f_{ij} 
- \sum_{\{j | (j,i) \in \mathcal{A}\}} f_{ji} =
\begin{cases}
R & \text{if $i = s$}, \\
-R & \text{if $i = t$}, \\
0 & \text{otherwise},
\end{cases}
\]
for all $i \in \mathcal{N}$, and
\[0 \le f_{ij} \le z_{ij}\]
for all $(i,j) \in \mathcal{A}$.
We assume, without loss of generality, that $f$ is cycle-free in the
sense that the subgraph 
$\mathcal{G}^\prime = (\mathcal{N}, \mathcal{A}^\prime)$, where
$\mathcal{A}^\prime := \{(i,j) \in \mathcal{A} | f_{ij} > 0\}$, 
is acyclic.  (If $\mathcal{G}^\prime$
has a cycle, then it can be eliminated by subtracting flow from $f$ 
around it.)

Using the conformal realization theorem
(see, for example, \cite[Section 1.1]{ber98}), we decompose $f$ into
a finite set of paths $\{p_1, p_2, \ldots, p_M\}$, 
each carrying positive 
flow $R_{m}$ for $m= 1, 2, \ldots, M$, such that
$\sum_{m=1}^M R_{m} = R$.  
We treat each path $p_m$ as a tandem network and use it to deliver
innovative packets at rate arbitrarily close to $R_m$, 
resulting in an overall rate 
for innovative packets arriving at node $t$
that is arbitrarily close to $R$.
A formal argument is in Appendix~\ref{app:formal_general_unicast}.

\subsubsection{Multicast connections}

The result for multicast connections 
is, in fact, a straightforward extension of that
for unicast connections.
In this case, rather than a single sink $t$, we have a set
of sinks $T$.  
As in the framework of static broadcasting (see \cite{shu03, shf00}), we
allow sink nodes to operate at different rates.
We suppose that sink $t \in T$ wishes to achieve rate
arbitrarily close to $R_t$, i.e.,\  to recover the $K$ message packets,
sink $t$ wishes to wait for a time $\Delta_t$ that is only marginally
greater than $K/R_t$.  
We further suppose that
\[
R_t \leq \min_{Q \in \mathcal{Q}(s,t)}
  \left\{\sum_{(i, j) \in \Gamma_+(Q)} z_{ij}
  \right\}
\]
for all $t \in T$.
Therefore, by the max-flow/min-cut theorem, there exists, for each
$t \in T$, a flow vector $f^{(t)}$ satisfying
\[
\sum_{\{j | (i,j) \in \mathcal{A}\}} f_{ij}^{(t)} 
- \sum_{\{j | (j,i) \in \mathcal{A}\}} f_{ji}^{(t)} =
\begin{cases}
R_t & \text{if $i = s$}, \\
-R_t & \text{if $i = t$}, \\
0 & \text{otherwise},
\end{cases}
\]
for all $i \in \mathcal{N}$, and
$f_{ij}^{(t)} \le z_{ij}$ for all $(i,j) \in \mathcal{A}$.  

For each flow vector $f^{(t)}$, we go through the same argument as that
for a unicast connection, and we find that the probability of error at
every sink node can be made arbitrarily small by taking $K$
sufficiently large.

We summarize our results regarding wireline networks with the following
theorem statement.

\begin{Thm}
Consider the lossy wireline packet network $(\mathcal{G}, z)$.
The random linear coding scheme described in
Section~\ref{sec:coding_scheme} is capacity-achieving for
multicast connections,
i.e.,\  for $K$ sufficiently large, it can achieve, with
arbitrarily small error probability, a multicast
connection 
from source node $s$ to sink nodes in the set $T$ at rate
arbitrarily close to $R_t$ packets per unit time for each $t \in T$ if
\[
R_t \leq \min_{Q \in \mathcal{Q}(s,t)}
  \left\{\sum_{(i, j) \in \Gamma_+(Q)} z_{ij}
  \right\}
\]
for all $t \in T$.\footnote{In 
earlier versions of this work \cite{lme04, lmk05-further}, we required the
field size $q$ of the coding scheme to approach infinity for
Theorem~\ref{thm:100} to hold.  This requirement is in fact not
necessary, and the formal arguments in 
Appendix~\ref{app:formal} do not require
it.}
\label{thm:100}
\end{Thm}

\noindent \emph{Remark.}  
The capacity region is determined solely by the average rate $z_{ij}$ at
which packets are received on each arc $(i,j)$.  Therefore, the packet
injection and loss processes, which give rise to the packet reception
processes, can take any distribution, exhibiting arbitrary correlations,
as long as these average rates exist.

\subsection{Wireless packet networks}

The wireless case is actually very similar to the wireline one.  The main
difference is that
we now deal with hypergraph flows rather than regular graph flows.

Suppose that we wish to establish a connection of rate arbitrarily close
to $R$ packets per unit time from source node $s$ to sink node $t$.
Suppose further that
\[
R \leq \min_{Q \in \mathcal{Q}(s,t)}
  \left\{\sum_{(i, J) \in \Gamma_+(Q)} \sum_{K \not\subset Q} z_{iJK}
  \right\},
\]
where $\mathcal{Q}(s,t)$ is the set of all cuts between $s$ and $t$, and
$\Gamma_+(Q)$ denotes the set of forward hyperarcs of the cut $Q$, i.e.\  
\[
\Gamma_+(Q) := \{(i, J) \in \mathcal{A} \,|\, i \in Q, J\setminus Q
\neq \emptyset\} .
\]
Therefore there exists a
flow vector $f$ satisfying
\[
\sum_{\{j | (i,J) \in \mathcal{A}\}} \sum_{j \in J} f_{iJj} 
- \sum_{\{j | (j,I) \in \mathcal{A}, i \in I\}} f_{jIi} =
\begin{cases}
R & \text{if $i = s$}, \\
-R & \text{if $i = t$}, \\
0 & \text{otherwise},
\end{cases}
\]
for all $i \in \mathcal{N}$, 
\begin{equation}
\sum_{j \in K} f_{iJj} \le \sum_{\{L \subset J | L \cap K \neq
\emptyset\}} z_{iJL}
\label{eqn:600}
\end{equation}
for all $(i,J) \in \mathcal{A}$ and $K \subset J$,
and $f_{iJj} \ge 0$
for all $(i,J) \in \mathcal{A}$ and $j \in J$.
We again decompose $f$ into a finite set of paths $\{p_1, p_2, \ldots,
p_M\}$, each carrying positive flow $R_m$ for $m = 1,2, \ldots, M$, such
that $\sum_{m=1}^M R_m = R$.  
Some care must be taken in the interpretation of the flow and its path
decomposition because, in a wireless transmission, the same packet may
be received by more than one node.
The details of the interpretation are in
Appendix~\ref{app:formal_wireless} and, with it,
we can use path $p_m$ to deliver
innovative packets at rate arbitrarily close to $R_m$, yielding the
following theorem.

\begin{Thm}
Consider the lossy wireless packet network $(\mathcal{H}, z)$.
The random linear coding scheme described in
Section~\ref{sec:coding_scheme} is capacity-achieving for
multicast connections,
i.e.,\  for $K$ sufficiently large, it can achieve, with
arbitrarily small error probability, a multicast
connection 
from source node $s$ to sink nodes in the set $T$ at rate
arbitrarily close to $R_t$ packets per unit time for each $t \in T$ if
\[
R_t \leq \min_{Q \in \mathcal{Q}(s,t)}
  \left\{\sum_{(i, J) \in \Gamma_+(Q)} \sum_{K \not\subset Q} z_{iJK}
  \right\}
\]
\label{thm:200}
\end{Thm}
for all $t \in T$.

\section{Error exponents for Poisson traffic with i.i.d.\  losses}
\label{sec:error_exponents}

We now look at the rate of decay of the probability of
error $p_e$ in the coding delay $\Delta$.  
In contrast to traditional error exponents where coding delay is
measured in symbols, we measure coding delay in time units---time
$\tau = \Delta$ is 
the time at which the sink nodes attempt to decode the
message packets.  The two methods of measuring delay are essentially
equivalent when packets arrive in regular, deterministic intervals.

We specialize to the case of Poisson traffic with i.i.d.\  losses.  
Hence, in the wireline case, the process $A_{ij}$ is a Poisson process
with rate $z_{ij}$ and, in the wireless case, the process $A_{iJK}$ is a
Poisson process with rate $z_{iJK}$.
Consider the unicast case for now, and
suppose we wish to establish a connection of rate $R$.
Let $C$ be the supremum of all asymptotically-achievable rates.

To derive exponentially-tight bounds on the probability of error, it is
easiest to consider the case where the links are in fact delay-free, and
the transformation, described in Section~\ref{sec:model}, for links with
delay has not be applied.  The results we derive do, however, apply in
the latter case.
We begin by deriving an upper bound on the probability of error.
To this end, we take a flow vector $f$ from $s$ to $t$ of size $C$
and, following the development in
Appendix~\ref{app:formal}, 
develop a queueing network from it that describes the propagation of
innovative packets for a given innovation order $\rho$.
This queueing network now becomes a Jackson network.
Moreover, as a consequence of Burke's
theorem (see, for example, \cite[Section 2.1]{kel79}) and the fact that
the queueing network is acyclic, the
arrival and departure processes at all stations are 
Poisson in steady-state.  

Let $\Psi_{t}(m)$ be the arrival time of the $m$th  
innovative packet at $t$, and let $C^\prime := (1-q^{-\rho})C$. 
When the queueing network is in steady-state, the arrival of innovative
packets at $t$ is described by a Poisson process of rate $C^\prime$.
Hence we have
\begin{equation}
\lim_{m \rightarrow \infty} \frac{1}{m}
\log \mathbb{E}[\exp(\theta \Psi_{t}(m))]
= \log \frac{C^\prime}{C^\prime - \theta} 
\label{eqn:1100}
\end{equation}
for $\theta < C^\prime$ \cite{bpt98, pal03}.
If an error occurs, then fewer than $\lceil R\Delta \rceil$
innovative packets are received by $t$ by 
time $\tau = \Delta$, which is
equivalent to
saying that $\Psi_{t}(\lceil R\Delta \rceil) > \Delta$.
Therefore,
\[
p_e \le \Pr(\Psi_{t}(\lceil R\Delta \rceil) > \Delta),
\]
and, using the Chernoff bound, we obtain
\[
p_e \le \min_{0 \le \theta < C^\prime}
\exp\left(
-\theta \Delta + \log \mathbb{E}[\exp(\theta \Psi_{t}(\lceil R\Delta
\rceil) )] 
\right) .
\]
Let $\varepsilon$ be a positive real number.  
Then using equation (\ref{eqn:1100}) we obtain, 
for $\Delta$ sufficiently large,
\[
\begin{split}
p_e &\le \min_{0 \le \theta < C^\prime}
\exp\left(-\theta \Delta
+ R \Delta \left\{\log \frac{C^\prime}{C^\prime-\theta} + \varepsilon \right\} \right)
\\
&= \exp( -\Delta(C^\prime-R-R\log(C^\prime/R)) + R\Delta \varepsilon) .
\end{split}
\]
Hence, we conclude that
\begin{equation}
\lim_{\Delta \rightarrow \infty} \frac{-\log p_e}{\Delta}
\ge C^\prime - R - R\log(C^\prime/R) .
\label{eqn:1110}
\end{equation}

For the lower bound, we examine 
a cut whose flow capacity is $C$.  We take one such cut and denote it by
$Q^*$.  It is
clear that, if fewer than $\lceil R\Delta \rceil$ distinct packets are
received across $Q^*$ in time $\tau = \Delta$, then an error occurs.
For both wireline and wireless networks, the arrival of
distinct packets across $Q^*$ is described by a Poisson
process of rate $C$.  
Thus we have
\[
\begin{split}
p_e &\ge \exp(-C\Delta)
\sum_{l = 0}^{\lceil R\Delta \rceil - 1}
\frac{(C\Delta)^l}{l!}  \\
&\ge \exp(-C \Delta)
\frac{(C\Delta)^{\lceil R\Delta \rceil -1}}
{\Gamma(\lceil R \Delta \rceil)} ,
\end{split}
\]
and, using Stirling's formula, we obtain 
\begin{equation}
\lim_{\Delta \rightarrow \infty} \frac{-\log p_e}{\Delta}
\le C - R - R\log(C/R) .
\label{eqn:1115}
\end{equation}

Since (\ref{eqn:1110}) holds for all positive integers $\rho$, we conclude from
(\ref{eqn:1110}) and (\ref{eqn:1115}) that
\begin{equation}
\lim_{\Delta \rightarrow \infty} \frac{-\log p_e}{\Delta}
= C - R - R\log(C/R) .
\label{eqn:1120}
\end{equation}

Equation (\ref{eqn:1120}) defines the asymptotic rate of decay of the
probability of error in the coding delay $\Delta$.  This asymptotic rate
of decay is determined entirely by $R$ and $C$.  Thus, for a packet
network with Poisson traffic and i.i.d.\  losses employing the coding
scheme described in Section~\ref{sec:coding_scheme}, 
the flow capacity $C$ of the minimum cut of the network is
essentially the sole figure of merit of importance in determining the
effectiveness of the coding scheme for large, but finite, coding delay.
Hence, in deciding how to inject packets to support the desired
connection, a sensible approach is to reduce our attention to this
figure of merit, which is indeed the approach taken in \cite{lrm06}.

Extending the result from unicast connections to multicast connections
is straightforward---we simply obtain (\ref{eqn:1120}) for each sink.

\section{Conclusion}

We have proposed a simple random linear coding scheme for reliable
communication over packet networks and demonstrated that it is
capacity-achieving 
as long as packets
received on a link arrive according to a process that has an average rate.
In the special case of
Poisson traffic with i.i.d.\  losses, we have given error exponents that
quantify the rate of decay of the probability of error with coding
delay.  Our analysis took into account various peculiarities of
packet-level coding that distinguish it from symbol-level coding.  Thus,
our work intersects both with information theory and networking theory
and, as such, draws upon results from the two usually-disparate fields
\cite{eph98}.  Whether our results have implications for particular
problems in either field remains to be explored.

Though we believe that the scheme may be practicable, we also believe
that, through a greater degree of design or use of feedback, the scheme
can be improved.  Indeed, feedback can be readily employed to reduce the
memory requirements of intermediate nodes by getting them to clear their
memories of information already known to their downstream neighbors.
Aside from the scheme's memory requirements, we may wish to improve its
coding and decoding complexity and its side information overhead.  We
may also wish to improve its delay---a very important performance factor
that we have not explicitly considered, largely owing to the difficulty
of doing so.  The margin for improvement is elucidated in part
in \cite{pfs05}, which analyses various packet-level coding schemes,
including ARQ and the scheme of this paper, and assesses their delay,
throughput, memory usage, and computational complexity for the two-link
tandem network of Figure~\ref{fig:two_links}.
In our search for such improved schemes, we may be aided
by the existing schemes that we have mentioned that apply to specific,
special networks.  

We should not, however, focus our attention solely on the packet-level
code.  The packet-level code and the symbol-level code collectively form
a type of concatenated code, and an endeavor to understand the
interaction of these two coding layers is worthwhile.  Some work
in this direction can be found in \cite{vem05}.

\section*{Acknowledgments}

The authors would like to thank Pramod Viswanath and John Tsitsiklis for
helpful discussions and suggestions.

\appendices

\section{Formal arguments for main result}
\label{app:formal}

Here, we give formal arguments for Theorems~\ref{thm:100}
and~\ref{thm:200}.  Appendices~\ref{app:formal_two_link_tandem},
\ref{app:formal_l_link_tandem}, and~\ref{app:formal_general_unicast}
give formal arguments for three special cases of Theorem~\ref{thm:100}:
the two-link tandem network, the $L$-link tandem network, and general
unicast connections, respectively.  Appendix~\ref{app:formal_wireless}
gives a formal argument for 
Theorem~\ref{thm:200} in
the case of general unicast connections.

\subsection{Two-link tandem network}
\label{app:formal_two_link_tandem}

We consider all packets received by node 2, namely $v_1,
v_2, \ldots, v_N$, to be innovative.  We associate with node 2
the set of vectors $U$, which varies with time and is initially empty,
i.e.\  $U(0) := \emptyset$.  If packet $x$ is received by node 2 at time
$\tau$, then its auxiliary encoding vector $\beta$ is added to $U$ at
time $\tau$,
i.e.\  $U(\tau^+) := \{\beta\} \cup U(\tau)$.

We associate with node 3 the set of vectors $W$, which again varies with
time and is initially empty.  Suppose that packet $x$, with auxiliary
encoding vector $\beta$, is received by node 3 at time $\tau$.  Let
$\rho$ be a positive integer, which we call the \emph{innovation order}.
Then we say $x$ is innovative if $\beta \notin \mathrm{span}(W(\tau))$
and $|U(\tau)| > |W(\tau)| + \rho - 1$.  If $x$ is innovative, then
$\beta$ is added to $W$ at time $\tau$.  

The definition of innovative is designed to satisfy two properties:
First, we require
that $W(\Delta)$, the set of vectors in $W$ when the scheme
terminates, is linearly independent.
Second, we require that, when a packet is received by node 3 and 
$|U(\tau)| > |W(\tau)| + \rho - 1$, 
it is innovative with high probability.  The
innovation order $\rho$ is an arbitrary factor that ensures that the
latter property is satisfied.

Suppose that
packet $x$, with
auxiliary encoding vector $\beta$, is received by node 3 at time $\tau$
and that $|U(\tau)| > |W(\tau)| + \rho - 1$.
Since $\beta$ is a random linear combination of vectors in $U(\tau)$, it
follows that $x$ is innovative with some non-trivial
probability.  More precisely, because $\beta$ is uniformly-distributed
over $q^{|U(\tau)|}$ possibilities, of which at least 
$q^{|U(\tau)|} - q^{|W(\tau)|}$ are not in $\mathrm{span}(W(\tau))$, it
follows that 
\[
\Pr(\beta \notin \mathrm{span}(W(\tau)))
\ge \frac{q^{|U(\tau)|}-q^{|W(\tau)|}}{q^{|U(\tau)|}}
= 1 - q^{|W(\tau)|-|U(\tau)|}
\ge 1 - q^{-\rho}.
\]
Hence $x$ is innovative with probability at least
$1-q^{-\rho}$.  Since we can always discard innovative packets, we
assume that the event occurs 
with probability exactly
$1-q^{-\rho}$.  If instead $|U(\tau)| \le |W(\tau)| + \rho - 1$, then we
see that $x$ cannot be innovative, and this remains true at
least until another arrival occurs at node 2.  Therefore, for an
innovation order of $\rho$, the
propagation of innovative packets through node 2 is
described by the propagation of jobs through a single-server queueing
station with queue size $(|U(\tau)| - |W(\tau)| - \rho + 1)^+$.

The queueing station is serviced with probability $1-q^{-\rho}$ whenever
the queue is non-empty and a received packet arrives on arc $(2,3)$.  We can
equivalently consider ``candidate'' packets that arrive with probability
$1-q^{-\rho}$ whenever a received packet arrives on arc $(2,3)$ and say that
the queueing station is serviced whenever the queue is non-empty and a
candidate packet arrives on arc $(2,3)$. 
We consider all packets received on arc $(1,2)$ to be candidate packets.

The system we wish to analyze, therefore, is the following simple
queueing system:  Jobs arrive at node 2 according to the arrival of
received packets on arc $(1,2)$ and, with the exception of the first
$\rho - 1$ jobs, enter node 2's queue.  
The jobs in node 2's queue are
serviced by the arrival of candidate packets on arc $(2,3)$ and exit 
after being serviced.  The number of jobs exiting is
a lower bound on the number of packets 
with linearly-independent auxiliary encoding vectors
received by node 3. 

We analyze the queueing system of interest using the fluid
approximation for discrete-flow networks (see, for example, \cite{chy01,
chm91}).
We do not explicitly account for the fact that the first $\rho-1$
jobs arriving at node 2 do not enter its queue because 
this fact has no effect on job throughput.
Let $B_1$, $B$, and $C$ be the counting processes for the arrival of
received packets on arc $(1,2)$, of innovative packets on
arc $(2,3)$, and of candidate packets on arc $(2,3)$,
respectively.
Let $Q(\tau)$ be the number of jobs queued for service at node 2 at
time $\tau$.
Hence $Q = B_1 - B$.  Let $X := B_1 - C$ and $Y := C - B$.  Then
\begin{equation}
Q = X + Y.
\label{eqn:100}
\end{equation}
Moreover, we have
\begin{gather}
Q(\tau) dY(\tau) = 0, \\
dY(\tau) \ge 0,
\end{gather}
and
\begin{equation}
Q(\tau) \ge 0
\end{equation}
for all $\tau \ge 0$, and
\begin{equation}
Y(0) = 0.
\label{eqn:110}
\end{equation}

We observe now that 
equations (\ref{eqn:100})--(\ref{eqn:110}) give us
the conditions for a Skorohod problem (see,
for example, \cite[Section 7.2]{chy01}) and, by the oblique reflection
mapping theorem, there is a well-defined, 
Lipschitz-continuous mapping $\Phi$ such that $Q = \Phi(X)$.  

Let
\begin{gather*}
\bar{C}^{(K)}(\tau) := \frac{C(K\tau)}{K}, \\
\bar{X}^{(K)}(\tau) := \frac{X(K\tau)}{K},
\end{gather*}
and
\[
\bar{Q}^{(K)}(\tau) := \frac{Q(K\tau)}{K}.
\]

Recall that $A_{23}$ is the counting process for the arrival of received
packets on arc $(2,3)$.  Therefore, $C(\tau)$ is the sum of
$A_{23}(\tau)$ Bernoulli-distributed random variables with parameter
$1-q^{-\rho}$.  
Hence
\[
\begin{split}
\bar{C}(\tau) 
&:= 
\lim_{K \rightarrow \infty}
\bar{C}^{(K)}(\tau) \\
& = 
\lim_{K \rightarrow \infty}
(1-q^{-\rho})
\frac{A_{23}(K \tau)}{K}
\qquad \text{a.s.} \\
&=
(1-q^{-\rho})z_{23} \tau
\qquad \text{a.s.},
\end{split}
\]
where the last equality follows by the assumptions of the model.
Therefore
\[
\bar{X}(\tau) :=
\lim_{K \rightarrow \infty} \bar{X}^{(K)}(\tau)
= (z_{12} - (1-q^{-\rho})z_{23}) \tau
\qquad \text{a.s.}
\]
By the Lipschitz-continuity of $\Phi$, then, it follows that
$\bar{Q} := \lim_{K \rightarrow \infty} \bar{Q}^{(K)} 
= \Phi(\bar{X})$,  i.e.\
$\bar{Q}$ is, almost surely, 
the unique $\bar{Q}$ that satisfies, for some
$\bar{Y}$,
\begin{gather}
\bar{Q}(\tau) =
(z_{12} - (1-q^{-\rho})z_{23})\tau + \bar{Y},
\label{eqn:200} \\
\bar{Q}(\tau) d\bar{Y}(\tau) = 0, \\
d\bar{Y}(\tau) \ge 0, 
\end{gather}
and
\begin{equation}
\bar{Q}(\tau) \ge 0
\end{equation}
for all $\tau \ge 0$, and
\begin{equation}
\bar{Y}(0) = 0.
\label{eqn:210}
\end{equation}

A pair $(\bar{Q}, \bar{Y})$ that satisfies
(\ref{eqn:200})--(\ref{eqn:210}) is 
\begin{equation}
\bar{Q}(\tau)
= (z_{12} - (1-q^{-\rho})z_{23})^+ \tau
\label{eqn:220}
\end{equation}
and
\[
\bar{Y}(\tau)
= (z_{12} - (1-q^{-\rho})z_{23})^- \tau.
\]
Hence $\bar{Q}$ is given by equation (\ref{eqn:220}).

Recall that node 3 can recover the message packets with high probability
if it receives $\lfloor K(1+\varepsilon) \rfloor$ packets with
linearly-independent auxiliary encoding vectors and that the number of
jobs exiting the queueing system is a lower bound on the number of
packets with linearly-independent auxiliary encoding vectors received by
node 3.  Therefore, node 3 can recover the message packets with high
probability if $\lfloor K(1 + \varepsilon) \rfloor$ or more jobs exit
the queueing system.  Let $\nu$ be the number of jobs that have exited
the queueing system by time $\Delta$.  Then
\[
\nu = B_1(\Delta) - Q(\Delta).
\]
Take $K = \lceil (1-q^{-\rho}) \Delta R_c R / (1 + \varepsilon) \rceil$, 
where $0 < R_c < 1$.  
Then
\[
\begin{split}
\lim_{K \rightarrow \infty} 
\frac{\nu}{\lfloor K(1 + \varepsilon) \rfloor}
&= \lim_{K \rightarrow \infty} 
\frac{B_1(\Delta) - Q(\Delta)}{K (1 + \varepsilon)} \\
&= \frac{z_{12} - (z_{12} - (1 - q^{-\rho}) z_{23})^+}
{(1 - q^{-\rho})R_cR} \\
&= \frac{\min(z_{12}, (1-q^{-\rho}) z_{23})}
{(1 - q^{-\rho})R_cR} \\
&\ge
\frac{1}{R_c} \frac{\min(z_{12}, z_{23})}{R} > 1
\end{split} 
\]
provided that
\begin{equation}
R \le \min(z_{12}, z_{23}).
\label{eqn:300}
\end{equation}
Hence, for all $R$ satisfying (\ref{eqn:300}), $\nu \ge \lfloor K(1 +
\varepsilon) \rfloor$ with probability arbitrarily close to 1 for $K$
sufficiently large.  The rate achieved is
\[
\frac{K}{\Delta} 
\ge \frac{(1-q^{-\rho}) R_c}{1 + \varepsilon} R,
\]
which can be made arbitrarily close to $R$ by varying $\rho$, $R_c$, and
$\varepsilon$.

\subsection{$L$-link tandem network}
\label{app:formal_l_link_tandem}

For $i = 2, 3, \ldots, L+1$, we associate with node $i$ the set of
vectors $V_i$, which varies with time and is initially empty. 
We define $U := V_2$ and $W := V_{L+1}$.
As in the case of the two-link tandem, 
all packets received by node 2 are considered
innovative and, if packet $x$ is received by
node 2 at time $\tau$, then its auxiliary encoding vector $\beta$ is
added to $U$ at time $\tau$.  
For $i = 3, 4, \ldots, L+1$,
if packet
$x$, with auxiliary encoding vector $\beta$, is received by node $i$ at
time $\tau$, then we say $x$ is innovative if 
$\beta \notin \mathrm{span}(V_i(\tau))$ and 
$|V_{i-1}(\tau)| > |V_{i}(\tau)| + \rho - 1$.
If $x$ is innovative, then $\beta$ is added to
$V_i$ at time $\tau$.  

This definition of innovative is a straightforward extension of that
in Appendix~\ref{app:formal_two_link_tandem}.  
The first property remains the same:
we continue to require
that $W(\Delta)$ is a set of linearly-independent vectors.
We extend the second property so that, when
a packet is received by node $i$ for any $i = 3, 4, \ldots, L+1$ and 
$|V_{i-1}(\tau)| > |V_{i}(\tau)| + \rho - 1$, 
it is innovative with high probability.  

Take some $i \in \{3, 4, \ldots, L+1\}$.  Suppose that packet $x$, with
auxiliary encoding vector $\beta$, is received by node $i$ at time
$\tau$ and that $|V_{i-1}(\tau)| > |V_i(\tau)| + \rho - 1$.  Thus, the
auxiliary encoding vector $\beta$ is a random linear combination of
vectors in some set $V_0$ that contains $V_{i-1}(\tau)$.  Hence, because
$\beta$ is uniformly-distributed over $q^{|V_0|}$ possibilities, of
which at least $q^{|V_0|} - q^{|V_i(\tau)|}$ are not in
$\mathrm{span}(V_i(\tau))$, it follows that
\[
\Pr(\beta \notin \mathrm{span}(V_i(\tau))) 
\ge \frac{q^{|V_0|} - q^{|V_i(\tau)|}}{q^{|V_0|}}
= 1 - q^{|V_i(\tau)| - |V_0|}
\ge 1 - q^{|V_i(\tau)| - |V_{i-1}(\tau)|}
\ge 1 - q^{-\rho} .
\]
Therefore $x$ is innovative with probability at least $1 - q^{-\rho}$.  

Following the argument in Appendix~\ref{app:formal_two_link_tandem}, we
see, for all $i = 2, 3, \ldots, L$, that
the propagation of innovative packets through node $i$ 
is described by the propagation of
jobs through a single-server queueing station with queue size
$(|V_i(\tau)| - |V_{i+1}(\tau)| - \rho + 1)^+$ and that the queueing
station is serviced with probability $1 - q^{-\rho}$ whenever the queue
is non-empty and a received packet arrives on arc $(i, i+1)$.
We again consider candidate packets that 
arrive with probability $1 - q^{-\rho}$ whenever a received packet
arrives on arc $(i, i+1)$ and say that the queueing station is serviced
whenever the queue is non-empty and a candidate packet arrives on arc
$(i, i+1)$.

The system we wish to analyze in this case is therefore the following
simple queueing network:  Jobs arrive at node 2 according to the arrival
of received packets on arc $(1,2)$ and, with the exception of the first
$\rho - 1$ jobs, enter node 2's queue.  For $i = 2, 3, \ldots, L - 1$,
the jobs in node $i$'s queue are serviced by the arrival of candidate
packets on arc $(i,i+1)$ and, with the exception of the first $\rho - 1$
jobs, enter node $(i+1)$'s queue after being serviced.  The jobs in node
$L$'s queue are serviced by the arrival of candidate packets on arc $(L,
L+1)$ and exit after being serviced.  The number of jobs exiting is a
lower bound on the number of packets with linearly-independent auxiliary
encoding vectors received by node $L+1$.

We again analyze the queueing network of interest using the fluid
approximation for discrete-flow networks, and we again do not explicitly
account for the fact that the first $\rho-1$ jobs arriving at a queueing
node do not enter its queue.  Let $B_1$ be the counting process for the 
arrival of received
packets on arc $(1,2)$.  For $i = 2, 3, \ldots, L$, let 
$B_i$, and $C_i$ be the counting processes for the arrival of 
innovative packets
and candidate packets on arc $(i, i+1)$, respectively.
Let $Q_i(\tau)$ be the number of jobs queued for service at node $i$ at
time $\tau$.  Hence, for $i = 2, 3, \ldots, L$, 
$Q_i = B_{i-1} - B_i$.  Let $X_i := C_{i-1} - C_i$
and $Y_i := C_i - B_i$, where $C_1 := B_1$.
Then, we obtain a Skorohod problem with the following conditions:
For all $i = 2, 3, \ldots, L$,
\[
Q_i = X_i - Y_{i-1} + Y_i.
\]
For all $\tau \ge 0$ and $i = 2, 3, \ldots, L$,
\begin{gather*}
Q_i(\tau) dY_i(\tau) = 0, \\
dY_i(\tau) \ge 0,
\end{gather*}
and
\begin{equation*}
Q_i(\tau) \ge 0.
\end{equation*}
For all $i = 2, 3, \ldots, L$, 
\begin{equation*}
Y_i(0) = 0.
\end{equation*}

Let
\[
\bar{Q}_i^{(K)}(\tau) := \frac{Q_i(K\tau)}{K}
\]
and $\bar{Q}_i := \lim_{K \rightarrow \infty} \bar{Q}^{(K)}_i$
for $i = 2, 3, \ldots, L$.
Then the vector $\bar{Q}$ 
is, almost surely, the unique $\bar{Q}$ that satisfies,
for some $\bar{Y}$, 
\begin{gather}
\bar{Q}_i(\tau) =
\begin{cases}
(z_{12} - (1-q^{-\rho})z_{23}) \tau + \bar{Y}_2(\tau) & \text{if $i=2$}, \\
(1-q^{-\rho})(z_{(i-1)i} - z_{i(i+1)}) \tau + \bar{Y}_i(\tau)
- \bar{Y}_{i-1}(\tau) & \text{otherwise},
\end{cases}
\label{eqn:400} \\
\bar{Q}_i(\tau) d\bar{Y}_i(\tau) = 0, \\
d\bar{Y}_i(\tau) \ge 0, 
\end{gather}
and
\begin{equation}
\bar{Q}_i(\tau) \ge 0
\end{equation}
for all $\tau \ge 0$ and $i = 2,3,\ldots, L$, and
\begin{equation}
\bar{Y}_i(0) = 0
\label{eqn:410}
\end{equation}
for all $i = 2,3,\ldots, L$.

A pair $(\bar{Q}, \bar{Y})$ that satisfies
(\ref{eqn:400})--(\ref{eqn:410}) is 
\begin{equation}
\bar{Q}_i(\tau)
= (\min(z_{12}, \min_{2 \le j < i} \{
(1-q^{-\rho})z_{j(j+1)}
\}) - (1-q^{-\rho}) z_{i(i+1)})^+ \tau
\label{eqn:420}
\end{equation}
and
\[
\bar{Y}_i(\tau)
= (\min(z_{12}, \min_{2 \le j < i} \{
(1-q^{-\rho})z_{j(j+1)}
\}) - (1-q^{-\rho}) z_{i(i+1)})^- \tau .
\]
Hence $\bar{Q}$ is given by equation (\ref{eqn:420}).  

The number of jobs that have exited the queueing network by time
$\Delta$ is given by
\[
\nu = B_1(\Delta) - \sum_{i=2}^L Q_i(\Delta).
\]
Take $K = \lceil (1-q^{-\rho}) \Delta R_c R / (1 + \varepsilon) \rceil$, 
where $0 < R_c < 1$.  
Then
\begin{equation}
\begin{split}
\lim_{K \rightarrow \infty} 
\frac{\nu}{\lfloor K(1 + \varepsilon) \rfloor}
&= \lim_{K \rightarrow \infty} 
\frac{B_1(\Delta) - \sum_{i=2}^LQ(\Delta)}{K (1 + \varepsilon)} \\
&=
\frac{\min(z_{12}, \min_{2 \le i \le L}\{(1-q^{-\rho}) z_{i(i+1)}\})}
{(1 - q^{-\rho})R_cR} \\
&\ge
\frac{1}{R_c} \frac{\min_{1 \le i \le L}\{z_{i(i+1)}\}}{R} > 1
\end{split} 
\label{eqn:490}
\end{equation}
provided that
\begin{equation}
R \le \min_{1 \le i \le L}\{z_{i(i+1)}\}.
\label{eqn:500}
\end{equation}
Hence, for all $R$ satisfying (\ref{eqn:500}), $\nu \ge \lfloor K(1 +
\varepsilon) \rfloor$ with probability arbitrarily close to 1 for $K$
sufficiently large.  The rate can again be made arbitrarily close to $R$
by varying $\rho$, $R_c$, and $\varepsilon$.

\subsection{General unicast connection}
\label{app:formal_general_unicast}

As described in Section~\ref{sec:wireline_unicast}, we decompose the
flow vector $f$ associated with a unicast connection into a finite set
of paths $\{p_1, p_2, \ldots, p_M\}$, each carrying positive flow $R_m$
for $m = 1,2,\ldots, M$ such that $\sum_{m=1}^M R_m = R$.  We now
rigorously show how each path $p_m$ can be treated as a separate tandem
network used to deliver innovative packets at rate arbitrarily close to
$R_m$.

Consider a single path $p_m$.  We write $p_m = \{i_1, i_2, \ldots,
i_{L_m}, i_{L_m+1}\}$, where $i_1 = s$ and $i_{L_m+1} = t$.  
For $l = 2, 3, \ldots, L_m+1$, we associate with node $i_l$ the set of
vectors $V^{(p_m)}_l$, which varies with time and is initially empty.
We define $U^{(p_m)} := V^{(p_m)}_2$ and $W^{(p_m)} :=
V^{(p_m)}_{L_m+1}$.
Suppose packet $x$, with auxiliary encoding vector $\beta$, is received by
node $i_2$ at time $\tau$.  We associate with $x$ the independent random
variable $P_x$, which takes the value $m$ with probability $R_m /
z_{si_2}$.  If $P_x = m$, then we say $x$ is innovative on path $p_m$,
and $\beta$ is added to $U^{(p_m)}$ at time
$\tau$.  Now suppose packet $x$, with auxiliary encoding vector $\beta$,
is received by node $i_l$ at time $\tau$, where $l \in \{3, 4, \ldots,
L_m+1\}$.  
We associate with $x$ the independent random
variable $P_x$, which takes the value $m$ with probability $R_m /
z_{i_{l-1}i_l}$.  We say $x$ is innovative on path $p_m$ if
$P_x = m$, $\beta \notin \mathrm{span}(V_l^{(p_m)}(\tau) \cup
\tilde{V}_{\setminus m})$,
and
$|V_{l-1}^{(p_m)}(\tau)| > |V_l^{(p_m)}(\tau)| + \rho - 1$, where
$\tilde{V}_{\setminus m} := \cup_{n=1}^{m-1} W^{(p_n)}(\Delta)
\cup \cup_{n=m+1}^M U^{(p_n)}(\Delta)$. 

This definition of innovative is somewhat more complicated than that in
Appendices~\ref{app:formal_two_link_tandem}
and~\ref{app:formal_l_link_tandem} because we now have $M$ paths that we
wish to analyze separately.  We have again designed the definition to
satisfy two properties:  First, we require that $\cup_{m=1}^M
W^{(p_m)}(\Delta)$ is linearly-independent.  This is easily verified:
Vectors are added to $W^{(p_1)}(\tau)$ only if they are linearly
independent of existing ones; vectors are added to $W^{(p_2)}(\tau)$
only if they are linearly independent of existing ones and ones in
$W^{(p_1)}(\Delta)$; and so on.  Second, we require that, when a packet
is received by node $i_l$, $P_x = m$, and $|V_{l-1}^{(p_m)}(\tau)| >
|V_l^{(p_m)}(\tau)| + \rho - 1$, it is innovative on path $p_m$ with
high probability.  

Take $l \in \{3, 4, \ldots, L_m+1\}$.  Suppose that packet $x$, with
auxiliary encoding vector $\beta$, is received by node $i_l$ at time
$\tau$, that $P_x = m$, and that $|V_{l-1}^{(p_m)}(\tau)| >
|V_l^{(p_m)}(\tau)| + \rho - 1$.  Thus, the auxiliary encoding vector
$\beta$ is a random linear combination of vectors in some set $V_0$ that
contains $V_{l-1}^{(p_m)}(\tau)$.  Hence $\beta$ is
uniformly-distributed over $q^{|V_0|}$ possibilities, of which at least
$q^{|V_0|} - q^{d}$ are not in $\mathrm{span}(V_l^{(p_m)}(\tau) \cup
\tilde{V}_{\setminus m}$), where $d := {\mathrm{dim}(\mathrm{span}(V_0)
\cap \mathrm{span}(V_l^{(p_m)}(\tau) \cup \tilde{V}_{\setminus m}))}$.
We have
\[
\begin{split}
d &= \mathrm{dim}(\mathrm{span}(V_0))
+ \mathrm{dim}(\mathrm{span}(V_l^{(p_m)}(\tau) \cup
\tilde{V}_{\setminus m}))
- \mathrm{dim}(\mathrm{span}(V_0 \cup V_l^{(p_m)}(\tau) \cup
  \tilde{V}_{\setminus m})) \\
&\le \mathrm{dim}(\mathrm{span}(V_0 \setminus V_{l-1}^{(p_m)}(\tau)))
+ \mathrm{dim}(\mathrm{span}(V_{l-1}^{(p_m)}(\tau)))
+ \mathrm{dim}(\mathrm{span}(V_l^{(p_m)}(\tau) \cup
\tilde{V}_{\setminus m})) \\
&\qquad
- \mathrm{dim}(\mathrm{span}(V_0 \cup V_l^{(p_m)}(\tau) \cup
  \tilde{V}_{\setminus m})) \\
&\le \mathrm{dim}(\mathrm{span}(V_0 \setminus V_{l-1}^{(p_m)}(\tau)))
+ \mathrm{dim}(\mathrm{span}(V_{l-1}^{(p_m)}(\tau)))
+ \mathrm{dim}(\mathrm{span}(V_l^{(p_m)}(\tau) \cup
\tilde{V}_{\setminus m})) \\
&\qquad
- \mathrm{dim}(\mathrm{span}(V_{l-1}^{(p_m)}(\tau) \cup V_l^{(p_m)}(\tau) \cup
  \tilde{V}_{\setminus m})) .
\end{split}
\]
Since $V_{l-1}^{(p_m)}(\tau) \cup \tilde{V}_{\setminus m}$
and $V_{l}^{(p_m)}(\tau) \cup \tilde{V}_{\setminus m}$
both form linearly-independent sets,
\begin{multline*}
\mathrm{dim}(\mathrm{span}(V_{l-1}^{(p_m)}(\tau)))
+ \mathrm{dim}(\mathrm{span}(V_l^{(p_m)}(\tau) \cup
\tilde{V}_{\setminus m})) \\
\begin{aligned}
&= \mathrm{dim}(\mathrm{span}(V_{l-1}^{(p_m)}(\tau)))
+ \mathrm{dim}(\mathrm{span}(V_l^{(p_m)}(\tau)))
+ \mathrm{dim}(\mathrm{span}(\tilde{V}_{\setminus m})) \\
&= \mathrm{dim}(\mathrm{span}(V_l^{(p_m)}(\tau)))
+ \mathrm{dim}(\mathrm{span}(V_{l-1}^{(p_m)}(\tau)
\cup \tilde{V}_{\setminus m})).
\end{aligned}
\end{multline*}
Hence it follows that
\[
\begin{split}
d &\le \mathrm{dim}(\mathrm{span}(V_0 \setminus V_{l-1}^{(p_m)}(\tau)))
+ \mathrm{dim}(\mathrm{span}(V_l^{(p_m)}(\tau))) 
+ \mathrm{dim}(\mathrm{span}(V_{l-1}^{(p_m)}(\tau)
\cup \tilde{V}_{\setminus m})) \\
&\qquad
- \mathrm{dim}(\mathrm{span}(V_{l-1}^{(p_m)}(\tau) \cup V_l^{(p_m)}(\tau) \cup
  \tilde{V}_{\setminus m})) \\
&\le \mathrm{dim}(\mathrm{span}(V_0 \setminus V_{l-1}^{(p_m)}(\tau)))
+ \mathrm{dim}(\mathrm{span}(V_l^{(p_m)}(\tau))) \\
&\le |V_0 \setminus V_{l-1}^{(p_m)}(\tau)|
+ |V_l^{(p_m)}(\tau)| \\
&= |V_0| - |V_{l-1}^{(p_m)}(\tau)|
+ |V_l^{(p_m)}(\tau)|,
\end{split}
\]
which yields
\[
d - |V_0| \le 
|V_l^{(p_m)}(\tau)|
- |V_{l-1}^{(p_m)}(\tau)|
\le -\rho.
\]
Therefore
\[
\Pr(\beta \notin \mathrm{span}(V_l^{(p_m)}(\tau) \cup
\tilde{V}_{\setminus m}))
\ge \frac{q^{|V_0|} - q^d}{q^{|V_0|}}
= 1 - q^{d - |V_0|} 
\ge 1 - q^{-\rho}.
\]

We see then that, if we consider only those packets such that $P_x = m$,
the conditions that govern the propagation of innovative packets are
exactly those of an $L_m$-link tandem network, which we dealt with in
Appendix~\ref{app:formal_l_link_tandem}.  By recalling the distribution
of $P_x$, it follows that the propagation of innovative packets along
path $p_m$ behaves like an $L_m$-link tandem network with average
arrival rate $R_m$ on every link.  Since we have assumed nothing special
about $m$, this statement applies for all $m = 1, 2, \ldots, M$.

Take $K = \lceil (1-q^{-\rho}) \Delta R_c R / (1 + \varepsilon) \rceil$, 
where $0 < R_c < 1$.  
Then, by equation (\ref{eqn:490}),
\[
\lim_{K \rightarrow \infty} 
\frac{|W^{(p_m)}(\Delta)|}{\lfloor K(1+\varepsilon) \rfloor}
> \frac{R_m}{R}.
\]
Hence
\[
\lim_{K \rightarrow \infty} 
\frac{|\cup_{m=1}^M W^{(p_m)}(\Delta)|}{\lfloor K(1+\varepsilon) \rfloor}
= \sum_{m=1}^M 
\frac{|W^{(p_m)}(\Delta)|}{\lfloor K(1+\varepsilon) \rfloor}
> \sum_{m=1}^M \frac{R_m}{R} = 1.
\]
As before, the rate can be made arbitrarily close to $R$ by varying
$\rho$, $R_c$, and $\varepsilon$. 

\subsection{Wireless packet networks}
\label{app:formal_wireless}

The constraint (\ref{eqn:600}) can also be written as
\[
f_{iJj} \le 
\sum_{\{L \subset J | j \in L\}} \alpha_{iJL}^{(j)} z_{iJL}
\]
for all $(i,J) \in \mathcal{A}$ and $j \in J$, where 
$\sum_{j \in L} \alpha_{iJL}^{(j)} = 1$ for all $(i,J) \in \mathcal{A}$ 
and $L \subset J$, and $\alpha_{iJL}^{(j)} \ge 0$ for all $(i,J) \in
\mathcal{A}$, $L \subset J$, and $j \in L$.
Suppose packet $x$ is placed on hyperarc
$(i,J)$ and received by $K \subset J$ at time $\tau$.  We associate with
$x$ the independent random variable $P_x$, which takes the value $m$
with probability 
$R_m \alpha_{iJK}^{(j)} / 
\sum_{\{L \subset J|j \in L\}} \alpha_{iJL}^{(j)} z_{iJL}$, where
$j$ is the outward neighbor of $i$ on $p_m$.  Using this definition of
$P_x$ in place of that used in Appendix~\ref{app:formal_general_unicast}
in the case of wireline packet networks, we find that the two cases
become identical, with the propagation of innovative packets
along each path $p_m$ behaving like a tandem network with average arrival
rate $R_m$ on every link.  

\bibliographystyle{IEEEtran}
\bibliography{IEEEabrv,inform_theory}

\end{document}